# Control over topological insulator photocurrents with light polarization

J. W. McIver[1,2]*, D. Hsieh[1]*, H. Steinberg[1], P. Jarillo-Herrero[1] & N. Gedik[1]

[1]*Department of Physics, Massachusetts Institute of Technology, Cambridge MA 02139.*

[2]*Department of Physics, Harvard University, Cambridge MA 02138.*

* These authors contributed equally to this work

Three-dimensional topological insulators[1-3] represent a new quantum phase of matter with spin-polarized surface states[4,5] that are protected from backscattering. The static electronic properties of these surface states have been comprehensively imaged by both photoemission[4-8] and tunneling[9,10] spectroscopies. Theorists have proposed that topological surface states can also exhibit novel electronic responses to light, such as topological quantum phase transitions[11-13] and spin-polarized electrical currents[14,15]. However, the effects of optically driving a topological insulator out of equilibrium have remained largely unexplored experimentally, and no photocurrents have been measured. Here we show that illuminating the topological insulator $Bi_2Se_3$ with circularly polarized light generates a photocurrent that originates from topological helical Dirac fermions, and that reversing the helicity of the light reverses the direction of the photocurrent. We also observe a photocurrent that is controlled by the linear polarization of light, and argue that it may also have a topological surface state origin. This approach may allow the probing of dynamic properties of topological insulators[11-15] and lead to novel opto-spintronic devices[16].

The surface electronic spectrum of the topological insulator $Bi_2Se_3$[6,17] has been shown to be characterized by a single helical Dirac dispersion[8] such that counter-propagating electrons carry opposite spin. Hence, pure spin currents, which are a net flow of spin without a net flow of charge, are expected to propagate along the surfaces of a topological insulator in equilibrium [Fig.1(a)]. It is theoretically believed that by optically driving a topological insulator out of equilibrium with circularly polarized light, these pure spin currents can be transformed into a spin-polarized net electrical current [Fig.1(b)][14,15]. The working principle is that circularly polarized light induces interband transitions with a probability that is sensitive to the surface state spin orientation[18,19], which is momentum (k) dependent. As a result, the surface states can be asymmetrically depopulated in *k*-space, which converts the pure spin currents from the Dirac cone into a net spin-polarized electrical current[15]. Because the bulk bands of $Bi_2Se_3$ are spin-degenerate, these photon helicity-dependent currents can only be induced on the surface. However, to date no photocurrents of any kind have been observed in any topological insulator. Isolating this helicity-dependent photocurrent requires that certain experimental challenges be addressed, including competing laser heating-induced thermoelectric currents and additional sources of surface and bulk photocurrents generated by other mechanisms.

In our experiment, 795 nm laser light was focused to a 100 μm spot size and the induced photocurrents ($j_y$) were measured [Fig.1(c)] across unbiased exfoliated $Bi_2Se_3$ devices[20] [Fig.1(d)]. Polarization dependent photocurrents were identified by measuring $j_y$ while rotating a λ/4 waveplate by an angle $\alpha$, which varied the laser polarization with a 180° period from linearly P-polarized in the scattering plane ($\alpha = 0°$), to left-circular ($\alpha = 45°$), to P ($\alpha = 90°$), to right-circular ($\alpha = 135°$), to P ($\alpha = 180°$).

Owing to the high thermoelectric power of $Bi_2Se_3$[21], laser induced heat gradients in the sample are expected to cause a bulk thermoelectric current background in addition to any photocurrents generated. To isolate the photocurrent response, we varied the heat gradient between the contacts by sweeping the laser spot position ($y$) across the $Bi_2Se_3$ device [Fig.1(e)] at a fixed polarization ($\alpha = 0°$). We find that a current develops that switches polarity across the sample and is finite exactly at the center of the sample ($y = 0$). The contribution to $j_y$ that switches polarity can be attributed to a thermoelectric current with electron-like carriers, which is consistent with our *n*-type native $Bi_2Se_3$ [see Supplementary Information (SI)]. On the other hand, the finite contribution to $j_y$ at $y = 0$, where the sample is evenly heated and the thermoelectric current should be minimal, can be attributed to a photocurrent that may encode aspects of the surface states' electronic response to light. Figure 1(f) shows that this current scales linearly with laser intensity, which is a characteristic feature of a photocurrent (SI). All subsequent measurements were performed at $y = 0$ and in this low laser intensity regime ($I < 60$ W/cm$^2$) (SI) where sample heating is minimized.

To investigate the role of spin in generating the photocurrent, we measured the light polarization dependence of $j_y$ at $y = 0$. Figure 2(a) shows that when light is obliquely incident in the *xz*-plane, $j_y$ exhibits a strong polarization dependence that is comprised of four components

$$j_y(\alpha) = C\sin 2\alpha + L_1 \sin 4\alpha + L_2 \cos 4\alpha + D \qquad (1)$$

The coefficient $C$ parameterizes a helicity-dependent photocurrent because rotating the $\lambda/4$ waveplate varies the light polarization between left- and right-circular with the functional form $\sin 2\alpha$. The helicity-dependence indicates that $C$ is generated through a spin-dependent process.

This is because left- and right-circularly polarized light preferentially interact with opposite spin polarizations that are either aligned or anti-aligned to the light's wavevector[18], depending on the helicity. The other coefficients in eq. (1) parameterize helicity-independent photocurrents that depend on the linear polarization of light ($L_1$ and $L_2$) and that are polarization-independent ($D$), which will be discussed later in the text.

We now move to understand if the spin-mediated photocurrent $C$ is generated by states in the helical Dirac cone. If this is the case, it should be possible to deduce the surface state spin distribution by comparing the magnitude of $C$ at different light angles of incidence. Because $C$ is generated transverse to the light scattering plane ($xz$-plane) in Fig.2(a), the opposing spin polarizations that are excited by the different helicities must have a spin component in the $xz$-plane and be asymmetrically distributed along the $y$-direction in $k$-space. Figure 2(b) shows that $C$ becomes very small when light is obliquely incident in the $yz$-plane, such that the device contacts lie in the light scattering plane. This indicates that the electrons involved in generating $C$ have a spin polarization that is locked perpendicular to their linear momentum. When light is normally incident, $C$ completely vanishes [Fig.2(c)], which is characteristic of an in-plane spin distribution but is more fundamentally required to vanish by the in-plane rotational symmetry of $Bi_2Se_3$[15]. Together these results reveal that the helicity-dependent photocurrent $C$ arises from the asymmetric optical excitation of the helical Dirac cone.

Having identified that $C$ arises from the Dirac cone, we seek to understand if the contributions $L_1$, $L_2$, and $D$ in Fig.2(a) also share this origin. In general, the interband transition probabilities that set photocurrent magnitudes can be highly temperature (T) dependent owing to the thermal broadening of the Fermi distribution and small changes in the electronic structure due to changes in the electron-phonon coupling strength[22]. Therefore, to understand if $L_1$, $L_2$, and $D$ are governed by the same interband transitions that give rise to $C$, we compare their detailed T dependence. The inset of Figure 3(a) shows that the fraction of incident photons absorbed by the sample, the absorptivity (see Methods Summary), exhibits a sharp decrease as T is raised from 15 K. This is generally consistent with the T dependences exhibited by $C$, $L_1$, $L_2$, and $D$ [Fig.3(a)]. However, there are two clearly distinct sets of behavior: $C$ and $L_1$ decrease monotonically to a constant and finite value between 60-293 K, whereas $D$ and $L_2$ decrease identically to zero after undergoing a polarity reversal between 60-200 K. The similar behavior shared by $L_1$ and $C$ strongly indicates that their generation mechanisms are deeply related and that $L_1$ may also have a Dirac cone origin. On the other hand, the $D$ and $L_2$ photocurrents likely share a different origin.

The origin of $D$ and $L_2$ is revealed through the photon polarization dependence of the absorptivity, which exhibits only a $\cos 4\alpha$ modulation [Fig.3(b)]. This is expected because the maxima of $\cos 4\alpha$ describe when the incident light is P-polarized, which is the polarization that is generally absorbed most strongly by solids[23]. The modulation amplitude is approximately 5% of the $\alpha$-independent background, which matches the percentage that the photocurrent component $L_2\cos 4\alpha$ modulates the $\alpha$-independent photocurrent $D$ [Fig.3(b)]. This observation, together with their identical temperature dependence [Fig.3(a)], shows that $L_2$ is a trivial modulation of the photocurrent $D$. Because the polarization dependence of the absorptivity is representative of the bulk index of refraction[23], this is an indication that the photocurrent represented by $D$ and $L_2$ likely has a bulk origin.

The observation of polarization-dependent photocurrents that stem from helical Dirac fermions ($C$ and $L_1$) coexisting with a bulk photocurrent ($D$ and $L_2$) in a topological insulator is novel and we elaborate on their possible microscopic mechanisms below. The photocurrents $C$ and $L_1$ arise through the asymmetric excitation of states in $k$-space and thus fall under the category of circular and linear photogalvanic effects respectively[24]. Circular and linear photogalvanic effects have similarly been observed together in spin-orbit coupled quantum well structures[24,25] where the Rashba spin-split valance and conduction bands provide the required asymmetric spin distribution. It has been theoretically shown for these systems that the two photogalvanic effects are linked and that their combined magnitude is a measure of the spin texture's trivial Berry's phase[26,27]. Photogalvanic currents have similarly been predicted to be a measure of the non-trivial Berry's phase in topological insulators[15]. However, determining the Berry's phase requires a quantitative measure of the Dirac cone contribution alone. This is challenging because the depopulation of the Dirac cone using high energy light necessarily implies a population of bulk-like excited states, which may also carry a net photogalvanic current [Fig4(a)]. Eliminating these contributions will be possible when more insulating samples become available and by extending these measurements into the lower energy (sub-bulk gap) THz radiation regime so that only inter-band transitions within the Dirac cone occur[15]. While Rashba spin-split quantum well states have been observed in the inversion layer of some $Bi_2Se_3$ samples, their relative contribution to the circular photogalvanic effect can be expected to be small [SI]. This is because the circular photogalvanic effect from Rashba spin-split bands will have an inherent cancellation effect arising from the presence of two oppositely spin-polarized Fermi surfaces, which is absent for topological surface states because of their single Fermi surface [SI]. While we have provided a physical understanding of the circular photogalvanic effect ($C$) in topological insulators, the linear photogalvanic effect ($L_1$) requires and awaits a more comprehensive theoretical treatment.

The bulk nature of the photocurrent described by $D$ and $L_2$ [Fig.3(b)] precludes a photogalvanic origin because the photogalvanic effect is only permitted at the surface of $Bi_2Se_3$ where spin-splitting is present in the electronic structure. This is therefore likely due to a different mechanism that is allowed in the bulk called the photon drag effect[28-31]. Photon drag describes photocurrents that result from the transfer of linear momentum from incident photons to excited carriers [Fig.4(b)], thus permitting a photocurrent even if states are symmetrically distributed in *k*-space. Helicity-independent photon drag photocurrents generated transverse to the direction of momentum transfer, consistent with what we observe, have been attributed in conventional semiconductors to an aspheric bulk band structure[30], which is also present in $Bi_2Se_3$ and may be the origin of *D* and $L_2$. Recently, a new helicity-dependent form of photon drag was observed alongside photogalvanic currents in a quantum well system[25]. It was proposed that the photon momentum transfer opened a spin-dependent relaxation channel in the spin-split valance band that created a spin-polarized current. A similar process may be able to take place on the surface of a topological insulator where the required spin-splitting is provided by the Dirac cone. However, the bulk spin-degeneracy of $Bi_2Se_3$ enables us to rule out this and related[32] bulk photocurrent contributions to *C* and $L_1$. While a photo-induced inverse spin Hall effect has been observed in GaAs and related materials, the exceptionally short spin lifetime of bulk optically spin oriented carriers will make contributions from this effect very small [SI].

Our measurements show that the polarization of light can be used to generate and control photocurrents originating from topological surface states. The photocurrents observed are only one of many possible non-equilibrium properties of a topologically ordered phase[11-15] and there are features in our data that call for a detailed theoretical treatment. In addition to the possibility of measuring fundamental physical quantities, like the Berry's phase[15,26,27], optically induced currents provide a promising route to generate and control spin-polarized currents purely at an isolated surface or buried interface, which could be harnessed for spintronic applications[16].

**Methods Summary.**

Bi$_2$Se$_3$ was synthesized and devices were fabricated using the techniques reported in Ref. 20. 80 fs pulses of 795 nm (1.56 eV) laser light were derived from a Ti:sapphire oscillator at a repetition rate of 80MHz. Data were corrected for small variations in laser intensity as a function of $\alpha$ due to λ/4 waveplate imperfections. A 50x microscope objective and a high resolution CCD camera were used to align the beam and device position with one μm accuracy. The absorptivity was determined by performing reflectivity measurements on a bulk single crystal from the same ingot used to fabricate devices.

**Acknowledgements.**

This work was supported by DOE award number DE-FG02-08ER46521, performed in part at the NSF funded Harvard Center for Nanoscale Systems and made use of the MRSEC Shared Experimental Facilities supported by the National Science Foundation under award number DMR - 0819762. J.W.M. acknowledges financial support from an NSF graduate research fellowship. D.H. acknowledges support from a Pappalardo postdoctoral fellowship. H.S. acknowledges support from the Israeli Ministry of Science. P.J-H. acknowledges support from DOE Early Career Award number DE.SC0006418 and Packard Fellowship.

**Fig. 1: Isolation of a photocurrent response from a thermoelectric current background.** **(a)** Pure spin currents from the Dirac cone in equilibrium due to cancelling electrical current contributions $j$. **(b)** Spin-polarized electrical current induced by optically driving the Dirac cone with circularly polarized light. **(c)** Schematic of the experimental geometry. The laser beam is incident on the device at the out-of-plane angle $\theta$ defined from the $xy$-plane and the in-plane angle $\phi$. Photon polarization varied by rotating the λ/4 waveplate (QWP) and photo-induced currents $j_y$ were measured. **(d)** AFM image of a typical two-terminal 120 nm thick $Bi_2Se_3$ device. **(e)** $j_y/I$ with light obliquely incident at $\theta = 56°$ in the $xz$-plane as a function of beam focus position $y$ at room temperature where $y = 0$ is the center of the sample. The red arrows in the inset represent beam position as it is scanned across the sample. **(f)** $j_y$ as a function of laser intensity $I$ at $y = 0$ at 15 K.

**Fig. 2: A surface photocurrent originating from helical Dirac fermions.** **(a)** $j_y(\alpha)/I$ with light obliquely incident at $\theta = 56°$ in the $xz$-plane at 15 K. The solid red line is a fit to eq. (1) and the fit results are shown. **(b)** $j_y(\alpha)/I$ with light obliquely incident at $\theta = 56°$ in the $yz$-plane at 15 K. **(c)** $j_y(\alpha)/I$ with light normally incident ($\theta = 90°$) and $\phi = 180°$ so that the laser electric field is perpendicular to the contacts at $\alpha = 0$ at 15 K.

**Fig. 3: Distinct photocurrent contributions separated by temperature dependence.** **(a)** Fit results for $j_y(\alpha)/I$ as a function of temperature for the geometry in Fig.2(a). Inset: Optical absorptivity as a function of temperature with P-polarized light ($\alpha = 0$) at $\theta = 56°$. **(b)** Percent change of the absorptivity and $L_2\cos4\alpha /D$ as a function of photon polarization at room temperature.

**Fig. 4: Microscopic mechanisms of photocurrent generation.** **(a)** $k$-space depiction of photon helicity-induced currents from the Dirac cone via the circular photogalvanic effect including contributions from excited states. **(b)** $k$-space depiction of photon drag. The optical transition arrows are tilted to account for the transfer of linear photon momentum.

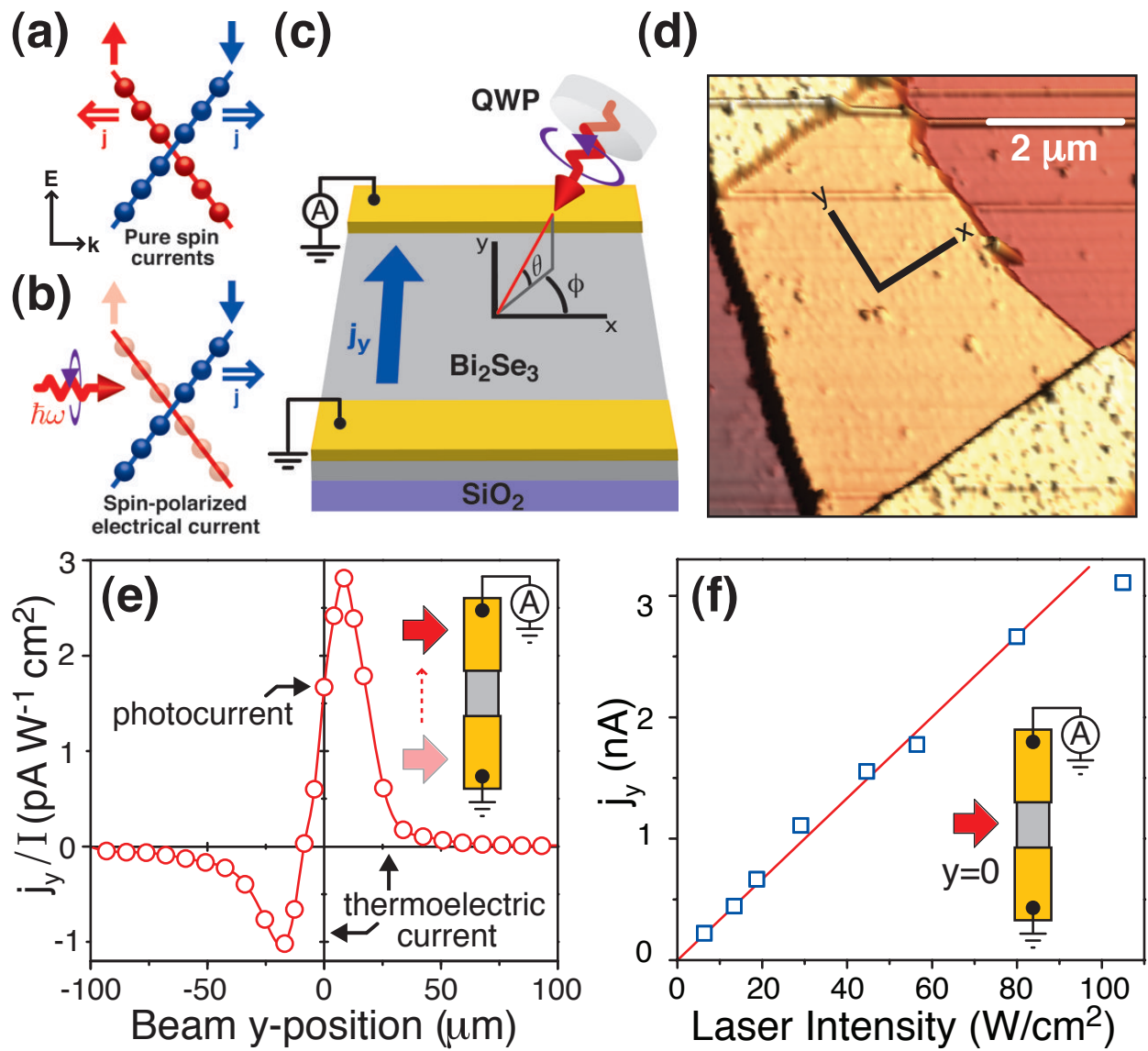

Fig.1 Gedik

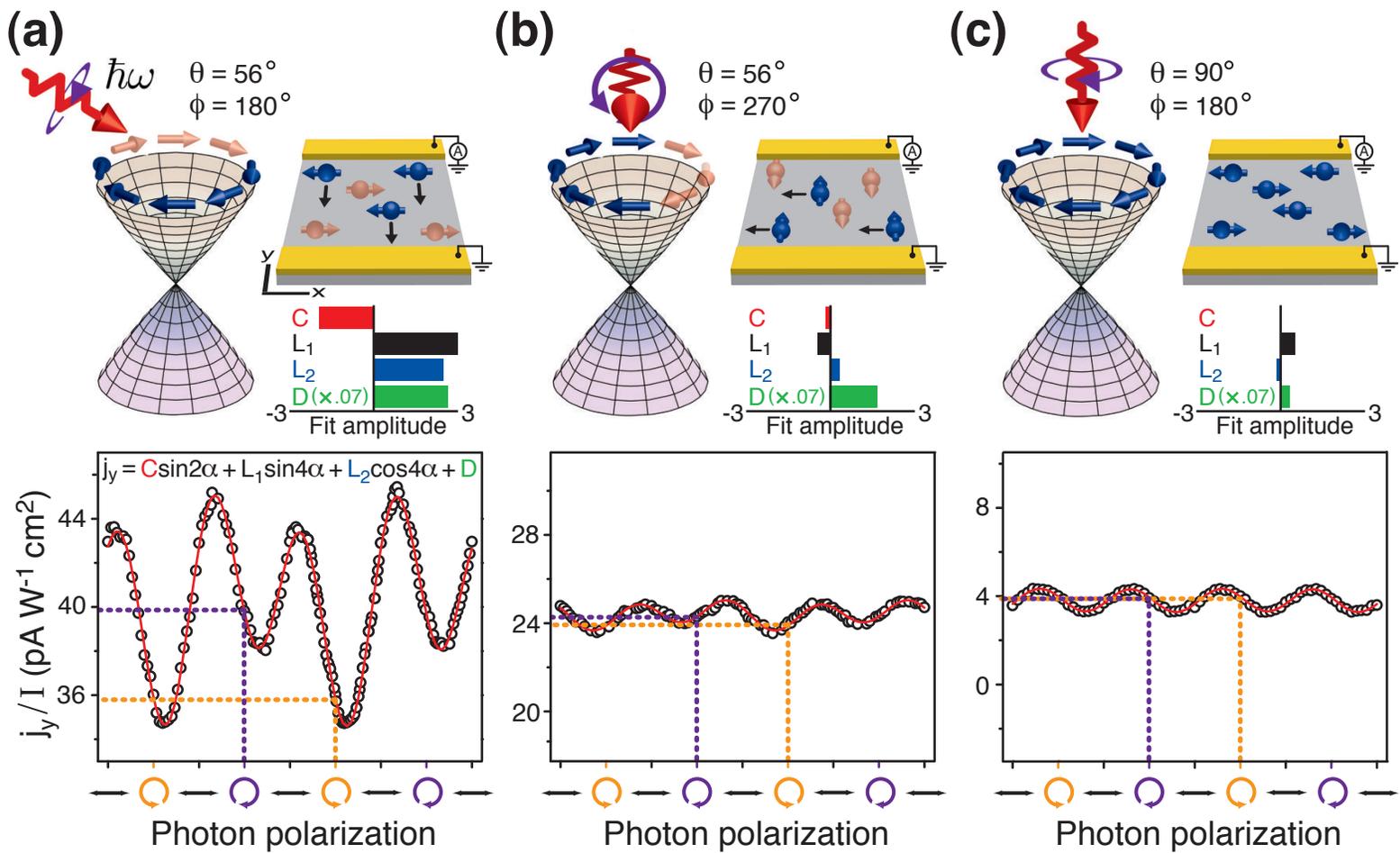

Fig.2 Gedik

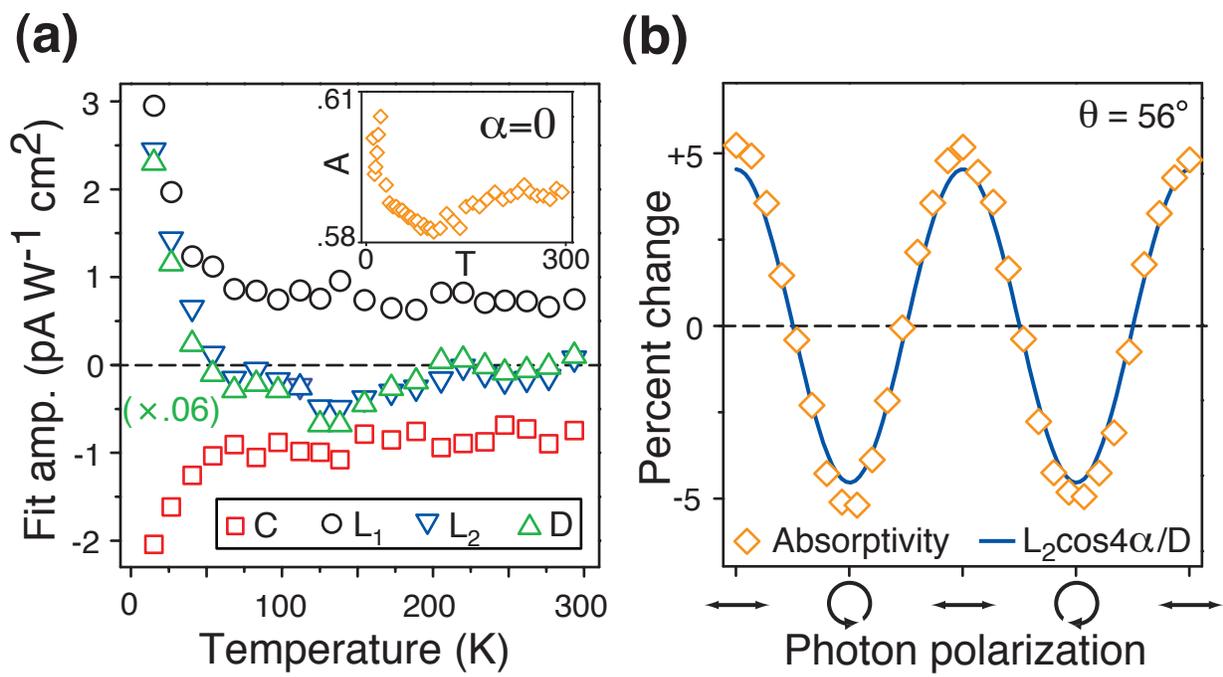

Fig.3 Gedik

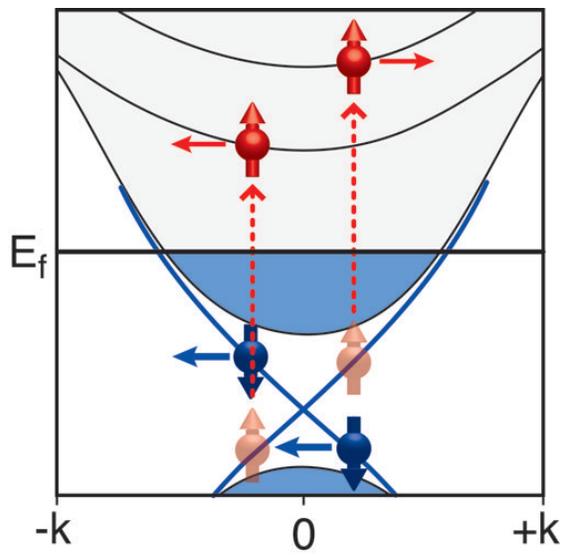 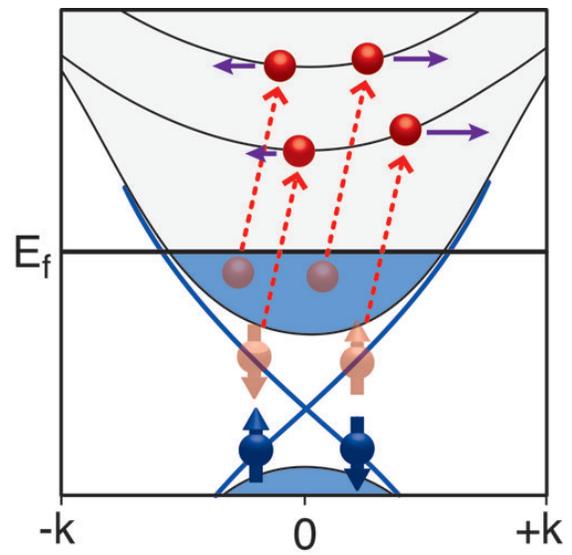

Fig.4 Gedik